# Voltage-triggered Ultra-fast Metal-insulator Transition in Vanadium Dioxide Switches

You Zhou, Xiaonan Chen, Changhyun Ko, Zheng Yang, Chandra Mouli and Shriram Ramanathan

*Abstract*— Electrically driven metal-insulator transition in vanadium dioxide ($VO_2$) is of interest in emerging memory devices, neural computation, and high speed electronics. We report on the fabrication of out-of-plane $VO_2$ metal-insulator-metal (MIM) structures and reproducible high-speed switching measurements in these two-terminal devices. We have observed a clear correlation between electrically-driven ON/OFF current ratio and thermally-induced resistance change during metal-insulator transition. It is also found that sharp metal-insulator transition could be triggered by external voltage pulses within 2 ns at room temperature and the achieved ON/OFF ratio is greater than two orders of magnitude with good endurance.

*Index Terms*—Metal-insulator transition, ultra-fast switch, correlated electrons, vanadium dioxide, memory.

## I. INTRODUCTION

VANADIUM dioxide ($VO_2$) is a strongly correlated electron system that undergoes a metal-insulator transition (MIT) in the vicinity of room temperature. Electrically triggered phase transition (E-MIT) in vanadium dioxide is of particular interest in novel devices for information storage, memory resistance and neural circuits [1-3]. An important parameter for such two terminal switches is the switching speed between the insulating and metallic state, which could provide information on both application potential and the transition mechanism. Recently, sub-nanosecond switching was demonstrated in $NbO_2$, a material showing thermally triggered MIT (T-MIT) at ~ 1080 K [4], whereas the reported switching speed of $VO_2$

Y. Zhou and S. Ramanathan are with the School of Engineering and Applied Sciences, Harvard University, Cambridge, MA 02138, USA (e-mail: youzhou@fas.harvard.edu; shriram@seas.harvard.edu).
X. Chen and C. Mouli are with Micron Technology Inc., Boise 83707, ID, USA (e-mail: xiaonanchen@micron.com; cmouli@micron.com).
C. Ko was at Harvard University, Cambridge, Massachusetts 02138, USA. He is now with the Department of Materials Science and Engineering, University of California, Berkeley, CA 94706 USA (e-mail: changhyun.ko@gmail.com).
Z. Yang was at Harvard University, Cambridge, MA 02138 USA. He is now with the Department of Electrical and Computer Engineering, University of Illinois at Chicago, Chicago, IL 60607 USA (e-mail: yangzhen@uic.edu).

devices is limited to several nanoseconds either due to the test structure or measurement setup limitations [5-8]. The demonstrated fastest switching speed (from electrical measurements) is ~5 ns for planar devices [5], and ~ 170 ns for out-of-plane devices [6]. Although out-of-plane metal-VO$_2$-metal structures are desirable for memory devices, direct growth on semiconducting substrates such as silicon limits the switching speed measurements due to the additional series resistance.

In this paper, we report on the growth and fabrication of metal/VO$_2$/metal structures by careful process control and study the switching properties of VO$_2$. A direct correlation between the transition magnitudes of E-MITs and those of T-MITs is observed and presented for the first time. We find that E-MITs triggered by a voltage pulse occur within less than 2 ns. The demonstrated ON/OFF resistance ratio is ~ 100, which is among the largest for pulse measurements to the best of our knowledge, for this material system.

## II. Experiments

VO$_2$ thin films were grown on various substrates under different conditions by rf-magnetron sputtering at 550ºC in an Ar/O$_2$ gas mixture from a V$_2$O$_5$ target (as summarized in Table I). Au/*c*-sapphire structures for pulsed measurements were prepared by depositing Ti (20 nm)/Au (200 nm) layers onto *c*-sapphire using electron-beam evaporation. Different oxygen partial pressures were used during sample growth to tune the oxygen to vanadium ratio and the film stoichiometry was studied by X-ray photoelectron spectroscopy with Al $K\alpha$ radiation (summarized in Tab. I). For electrical characterization of samples A to E, 300 μm by 300 μm Ti/Au square electrodes deposited by electron-beam evaporation were used as top electrodes, while the conducting substrates were bottom electrodes as in Fig. 1(a). For sample F, a tungsten probe (Tungsten, tip radius ~ 15 μm) was used to contact the Ti/Au layer that was used as bottom electrode, while a super-thin/soft probe needle (Tungsten, tip radius from ~ 0.1 μm to 0.5 μm) contacted VO$_2$ film as a top electrode as shown in Fig. 1(b). Keithley 2635A and HP 4156C were used for DC measurements. For pulse measurement, a pulse generator Agilent 81110A was used as source and HP Infinium oscilloscope 54832B measured the voltage on the sensing resistance $R_S$ as shown in Fig. 1(c). The current passing through the device is given by $V_R$ divided by $R_S$. The voltage was applied on the bottom electrode so that lower parasitic capacitance and less voltage variation from the top electrode can benefit the bandwidth of the sensing circuit.

TABLE I
METAL-INSULATOR TRANSITION PROPERTIES OF DIFFERENT SAMPLES

| Sample | Substrate | O$_2$/Ar partial pressure (mTorr/mTorr) | O/V ratio | Film Stoichiometry | Thickness (nm) | Device Size | $M_T$ ($\Omega/\Omega$) | $M_E$ ($\Omega/\Omega$) |
|---|---|---|---|---|---|---|---|---|
| A | Al (100) | 0/20 | ~2.1 | Stoichiometric | ~370 | 300 μm×300 μm | ~2855 | 146±10 |
| B | n$^+$-Si (100) | 0/20 | ~2.2 | Stoichiometric | ~370 | 300 μm×300 μm | ~2784 | 129±47 |
| C | n$^+$-Si (100) | 0.03/9.96 | ~2.1 | Stoichiometric | ~300 | 300 μm×300 μm | ~670 | 82±9 |
| D | n$^+$-Si (100) | 0/10 | ~1.9 | O$_2$ deficient | ~300 | 300 μm×300 μm | ~50 | 55±20 |
| E | n$^+$-Si (100) | 0.14/9.87 | ~2.3 | O$_2$ rich | ~200 | 300 μm×300 μm | ~15 | 11±7 |
| F | Au/Al$_2$O$_3$ (001) | 0.03/9.96 | ~2.1 | Stoichiometry | ~400 | 0.1 to 0.5 um radius | ~500 | ~150 |

## III. RESULTS AND DISCUSSION

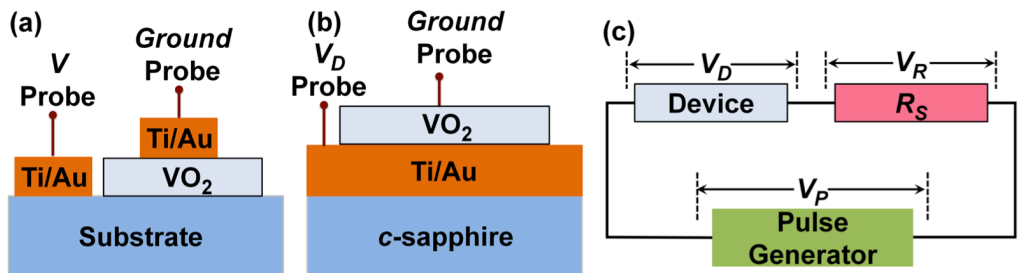

Fig. 1. Schematics of fabricated VO$_2$ out-of-plane devices (a) on samples A-E and (b) on sample F. (c) The voltage pulse is applied by a voltage pulse generator and the current through the device is monitored by the voltage drop on a sensing resistance R$_S$.

Fig. 2(a) shows the representative photoelectron spectra taken from samples C, D and E. The O to V ratio (summarized in Tab. I) is estimated by the vanadium 2p$_{3/2}$ peak positions and the integrated area of the O 1s and V 2p$_{3/2}$ peaks taking into account the relative sensitivity factors for O 1s and V 2p$_{3/2}$ respectively. Fig. 2(b) shows the normalized resistance, $R_N = R(T)/R(25\ ºC)$, versus temperature curves measured from the various samples. Fig. 2(c) shows the representative current versus electrical field curves from out-of-plane two terminal devices on different samples measured at room temperature. Current jump caused by E-MIT could be observed in all the samples and there is no obvious voltage polarity dependence. The magnitude of E-MIT, $M_E$, is defined by the resistance values at insulating state divided by that of the metallic phase. The magnitude of T-MIT, $M_T$, is given by the ratio of the resistance at 25 ºC to that at 100 ºC. Table I shows the data on $M_E$ taken from nearly ten different devices on each sample. From Fig. 2(c) and Table I, a trend emerges for samples A to E that the sample with a larger $M_T$ (near-stoichiometric) also has a larger $M_E$. The clear correlation between $M_E$ and $M_T$ serves as strong evidence that the VO$_2$ threshold switching phenomenon is a result of the bulk phase transition. In materials where the resistive switching is based on the rupture and formation of conducting filaments (CF) composed of defects, resistive switching behaviors are often observed in non-stoichiometric samples. Examples include ZnO$_x$, NiO$_x$ and other oxides [9-11]. This is because the oxygen vacancies (e.g. ZnO$_x$) or the cation defects (e.g. NiO$_x$) serve as the conducting paths in a higher resistance matrix [9, 10]. This appears to not be the case here wherein stoichiometric samples have the largest $M_E$. It is also found that for devices on different samples, the threshold field ranges from $10^6 – 10^7$ V/m, in agreement with previous reports [5]. In sample F $M_E$ is close to $M_T$, suggesting that the E-MITs happen close to bulk form under the electrode. Note that $M_E$ in other samples is ~1 order of magnitude smaller than $M_T$, which is caused by an interfacial oxide layer at VO$_2$/Al and VO$_2$/Si interfaces. Fig. 2(d) shows the resistance of the ON and OFF states as a function of device size measured from sample F. Both resistance values scale with the device area, which again indicates that E-MIT happens in the bulk of the film. For devices larger than 1 μm$^2$, the data become less meaningful because of the low resistance of the device that is comparable to the series resistance.

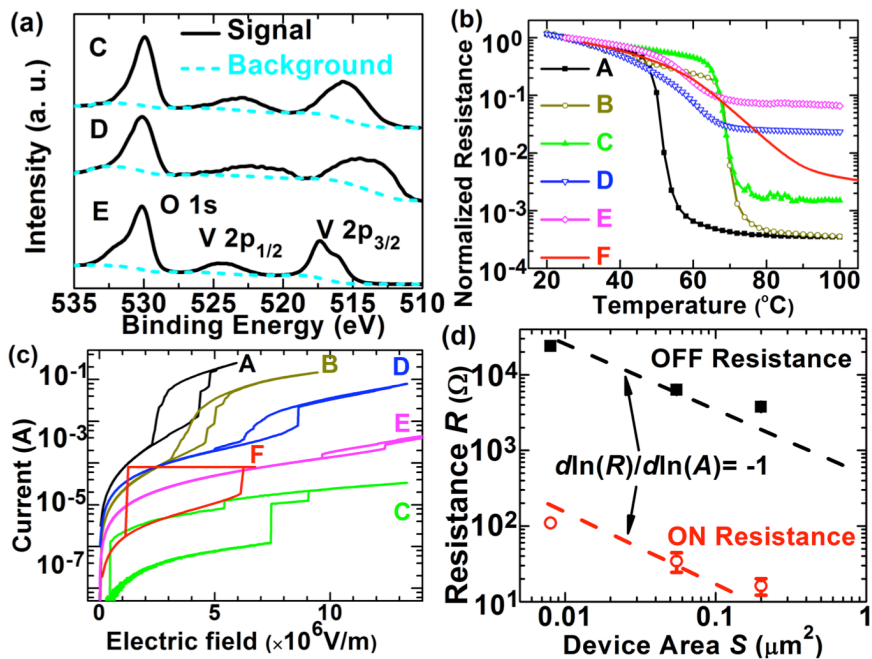

Fig. 2. (a) Representative X-ray photoelectron spectra taken from samples from C, D and E to study the sample stoichiometry. (b) Normalized resistance vs. temperature curves from the various samples listed in Table I. (c) Representative threshold switching phenomena in all samples. (d) The ON and OFF state resistance as a function of device size on sample F.

Fig. 3(a) shows voltage pulse source, $V_P$ and device current, $I_D$ as a function of time taken from a device (0.1 μm tip radius) on sample F. At a threshold voltage of ~ 2.25 V, the device is switched ON sharply with a very short switching time. The current overshoot is due to displacement current. $V_{IMT}$ in pulse measurement is almost the same with the value in DC measurements. The ON state resistance is 1.1 × 10$^2$ Ω. In the OFF state, the current is too small to be detected by an oscilloscope but we can estimate it to be ~ 2.4 × 10$^4$ Ω from DC measurements. Therefore, the transition magnitude is roughly 2 × 10$^2$, which is among the largest reported (~ 10 times in ref [5, 6, 12]). Note that the voltage pulse triggered transition has a similar ON/OFF ratio with the T-MIT magnitude, consistent with the observations in Fig. 2(d). Fig. 3(b) shows the magnified rising edge of E-MIT. The current first increases abruptly at the threshold and starts to oscillate due to series resistance and capacitance. Within 2 ns, the device resistance changes from ~ 2.4 × 10$^4$ Ω to 150 Ω. Here we define the switching time $\tau$ as the time needed for the device resistance to change from $R_{OFF}$ to $R_{OFF}(R_{ON}/R_{OFF})^{0.9}$ (i.e. resistance change by 90% in log scale). The switching time $\tau$ of the device in Fig. 3(b) is ~ 1.9 ns, much shorter than that reported out-of-plane devices on n$^+$-Si. Recent

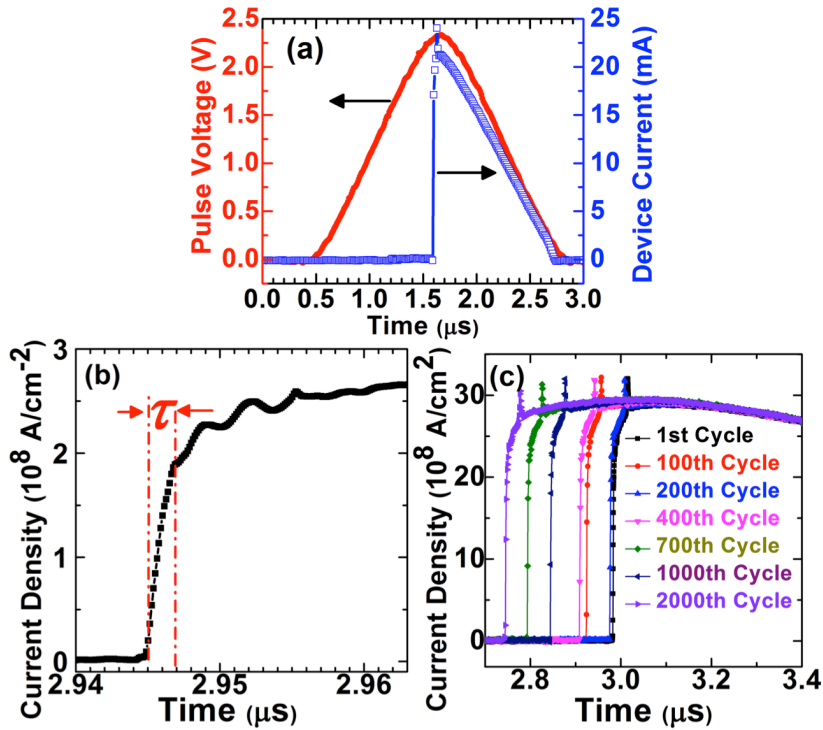

Fig. 3 (a) The external voltage pulse and current through the VO$_2$ MIM device, $I_D$, in a pulse measurement. (b) Magnified device current density versus time curve showing switching time of ~ 2 ns. (c) Multiple cycles of voltage pulse triggered metal-insulator transition. No obvious degradation in ON/OFF resistance and switching time is observed.

simulations based on a thermal circuit model suggest that the switching time of a VO$_2$ out-of-plane device (100 nm by 100 nm wide, 100 nm thick VO$_2$ device on Si) would be larger than 6 ns if current-induced Joule heating is the primary mechanism [13]. The switching time should be larger than 10 ns for the 400 nm thick films studied here, as it would scale correspondingly. Therefore, simply the Joule heating model alone may not explain the ultrafast E-MIT. Recently it has been demonstrated that E-MIT could be triggered by an electric field without passing current, which is attributed to electronic correlation effects [14]. In the present study, carrier injection which happens at a faster time scale than thermal heating could lead to an insulator-to-metal transition and explain the dynamics. Once the device is switched ON, the current conduction increases which could help to stabilize bulk MIT. Fig. 3(c) shows the switching behavior of a specific device after multiple pulsed cycles. There is negligible change in the ON state resistance as evidenced by the shape of the current pulse. The switching time $\tau$ of all the E-MITs shown is within ~2 ns. We note there that the transition timescale to recover the original state is also an important parameter and could be comparable in nanoscale devices owing in part to the increased thermal conductivity in the metallic state [13].

## IV. CONCLUSION

We have shown a clear correlation between E-MIT magnitude and sample stoichiometry in VO$_2$. We show that E-MITs happen close to a bulk form, being different from other CF-based resistive switching phenomena. The growth and fabrication of high quality VO$_2$ MIM two terminal devices enabled us to obtain sharp E-MITs with large ON/OFF ratio and probe the intrinsic switching in VO$_2$. The fast switching coupled with room temperature compatibility could lead to further interest in the use of correlated oxide phase transition devices for future electronics.